# A correction to collision rates of droplets in turbulent flows

Huang Zhang[*1], Qianfeng Liu[2]

1. Department of Thermal Engineering, Tsinghua University, Beijing, China, 100084;
2. Institute of Nuclear and New Energy Technology, Tsinghua University, Beijing, China, 100084

Corresponding Author*: zhanghuang15@mail.tsinghua.edu.cn

**Abstract**

This paper makes a correction to the collision rates of small droplets in turbulent fluid derived by Saffman and Turner(1956). The droplets are considered to be much smaller than the Kolmogorov length scale of turbulence and so the collision rates depend on the concentrations, dimensions of droplets, and the local structure of fluid. Not only the distortion but also the rotation of the fluid is taken into account between two close droplets. A rotation reference is fixed on one drop, and the fluxes of the other drops moving towards the fixed one are carried out based on this new reference. The sketch figure of fluxes is performed, and in three dimension space, the flux of the same drops consists a group of saddle surfaces, instead of hyperbolic curves in two dimension. The behaviors of turbulent flow are analyzed within the smallest eddies under the rotation reference, and a correction is made to the collision rates by multiplying a factor $\sqrt{2}$. The new rates are closer than the Saffman and Turner's to the ones obtained by DNS results.

## 1. Introduction

Droplet collision phenomena are common in our daily life (Mashall and Li, 2014). The formation of cloud is thought to be due to droplet collision induced by turbulence flow in the atmosphere of the earth (Grabowski and Wang, 2013). And in many industrial applications, such as spray cooling and combustion, droplet collision is often encountered. Especially in the moisture separator of steam generator of PWR, droplet collision is a key factor to influence the pressure drop and separation efficiency of the separator (Zhang et al., 2015).

In spite of the process of droplet collision is very complex, it is easier to study the collision rates of large number of droplets than the details of two or three droplets collision mutually. With the developing of computer modeling, direct numerical simulation (DNS) seems a powerful way to get the collision rates in tracing numerous droplets in turbulence (Sundaram and Collins, 1997; Zhou et al., 2001). However, DNS could not show the physical meaning of its results. Hence, on the other hand, Abrahamson (1975) derived a theoretical model of the collision rates of particles with large inertia in high energy flows. Whereas, small inertial particles are crucial in most situations. In the pioneer work of Saffman and Turner (1956), a theory of collision between small drops in a turbulent fluid was proposed. The collision rate of drops that are smaller than the small eddies of turbulence depends only on the dimensions of the drops, the rate of energy dissipation $\epsilon$ and the kinematic viscosity $\nu$. Their derivation was based on a main assumption that the inertia of drop in the small eddies of turbulence, which are also called Kolmogorov length, is so small that could be omitted. As a result, the drops are moving with the fluid with no relative motion. Then the mechanism of collision between drops is attributed to local shear motion of fluid. In addition, they thought, for two close points in a turbulent fluid, the relative motion is that of uniform strain.

However, the character of the motion in the neighborhood of any point of fluid is well-defined in any classical text book of fluid mechanics (Batchelor, 1967). Suppose the velocity of the fluid at position **x** ant time $t$ is **u(x,**$t$**)**, and the simultaneous velocity at a neighboring position **x+r** is **u(x+r,**$t$**)**. Where, for rectangular co-ordinates,

$$\delta \mathbf{u} = \mathbf{u}(\mathbf{x} + \mathbf{r}, t) - \mathbf{u}(\mathbf{x}, t) = \mathbf{r} \cdot \nabla \mathbf{u}(\mathbf{x}, t) \tag{1}$$

correct to the first order in the small distance **r** between the two points. For convenience, Eq. (1) is rewritten in the component form of tensor.

$$\delta u_i = r_j \frac{\partial u_i}{\partial x_j} \tag{2}$$

The geometrical character of the relative velocity $\delta \mathbf{u}$ can be recognized by decomposing $\frac{\partial u_i}{\partial x_j}$, which is a second-order tensor, into parts which are symmetrical and anti-symmetrical in the suffices $i$ and $j$. Thus we write $\delta u_i = \delta u_i^{(s)} + \delta u_i^{(a)}$, where

$$\delta u_i^{(s)} = r_j e_{ij}, \quad \delta u_i^{(a)} = r_j \xi_{ij}, \tag{3}$$

and

$$e_{ij} = \frac{1}{2}\left(\frac{\partial u_i}{\partial x_j} + \frac{\partial u_j}{\partial x_i}\right), \quad \xi_{ij} = \frac{1}{2}\left(\frac{\partial u_i}{\partial x_j} - \frac{\partial u_j}{\partial x_i}\right). \tag{4}$$

The pure straining motion of two close points is characterized by the rate-of-strain tensor $e_{ij}$. And we see that $\xi_{ij}$ is an anti-symmetrical tensor with only three independent components and may quite generally be written in the form

$$\xi_{ij} = -\frac{1}{2}\varepsilon_{ijk}\Omega_k \tag{5}$$

$\Omega_k$ is the k-component of the local vorticity $\boldsymbol{\Omega}(\mathbf{x},t)$, which is equal to $\nabla \times \mathbf{u}(\mathbf{x},t)$. In Saffman and Turner's paper, they regarded that the relative motion near two close points is only depend on $e_{ij}$, whereas, $\xi_{ij}$ should not be neglected since $\boldsymbol{\Omega}$ is always unequal to zero in some place in turbulent fluid. So we must make up the Saffman and Turner's work in a more reasonable way.

**2. Theoretical derivation**

Firstly, we choose a moving reference (*xyz*) fixed at a drop located at position $\mathbf{x}'$ with velocity $\mathbf{u}'(\mathbf{x}',t)$ and vorticity $\boldsymbol{\Omega}'(\mathbf{x}',t) = \nabla \times \mathbf{u}'(\mathbf{x}',t)$, seeing in figure 1. The directions of the moving co-ordinates are rearranged to lead to the non-diagonal elements of $e_{ij}$ are zero in this moving reference, which is always possible, for $e_{ij}$ is a symmetrical second-order tensor. So the Saffman and Turner's assumption that the relative motion of two close points is that of uniform strain is reasonable under this condition. For two neighboring points, where one of them is the origin **O**, the velocity of another point in the *xyz* co-ordinate is

$$w_i = x_i e_{ij},\tag{6}$$

where $e_{ij} = \begin{cases} e_{11} \text{ or } e_{22} \text{ or } e_{33}, & \text{if } i = j = 1,2,3 \\ 0, & \text{if } i \neq j \end{cases}$.

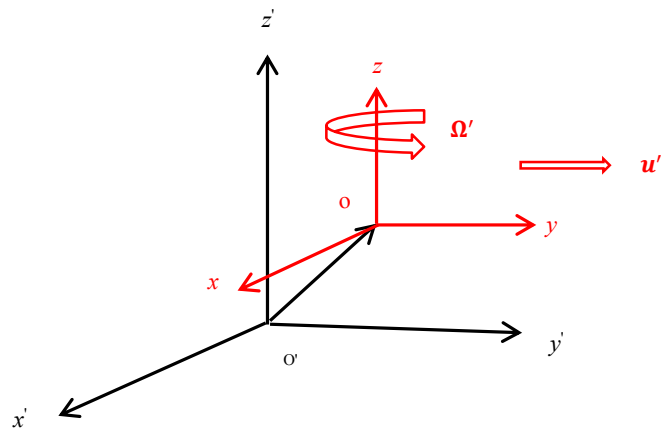

Figure 1. The fixed and the moving frame

We take a material element that is at position $\mathbf{x_0}$ near the origin at time $t=0$. After a given time $t$, the material element moves to another position $\mathbf{x}$. According to Eq. (6), the relationship between the position and the velocity of the material element is as

$$\frac{d\mathbf{x}}{dt} = \mathbf{w},\tag{7}$$

where the components of the equation above are as below

$$\begin{cases} \frac{dx_1}{dt} = w_1 \\ \frac{dx_2}{dt} = w_2 \\ \frac{dx_3}{dt} = w_3 \end{cases}.\tag{8}$$

When applying Eq. (6) to solve Eq. (8), we find that

$$\begin{cases} \frac{dx_1}{dt} = x_1 e_{11} \\ \frac{dx_2}{dt} = x_2 e_{22} \\ \frac{dx_3}{dt} = x_3 e_{33} \end{cases} \Rightarrow \begin{cases} \frac{dx_1}{x_1} = e_{11} dt \\ \frac{dx_2}{x_2} = e_{22} dt \\ \frac{dx_3}{x_3} = e_{33} dt \end{cases} \Rightarrow \begin{cases} x_1 = x_{10} e^{e_{11} t} \\ x_2 = x_{20} e^{e_{22} t} \\ x_3 = x_{30} e^{e_{33} t} \end{cases}\tag{9}$$

associated with the fact that $e_{11} = \left.\frac{\partial w_1}{\partial x_1}\right|_{\mathbf{x}=0}$, $e_{22} = \left.\frac{\partial w_2}{\partial x_2}\right|_{\mathbf{x}=0}$ and $e_{33} = \left.\frac{\partial w_3}{\partial x_3}\right|_{\mathbf{x}=0}$ are all constants. Special attention should be paid here that, although the velocity of the origin $\mathbf{O}$ is zero, the variation rate of the velocity along the axes may not be zero.

We prefer to acknowledging the relation of $x_1$, $x_2$ and $x_3$ to get the trajectory of the material element. The key procedure is to eliminate the variable $t$ in Eq. (8). We multiply the components of position $\mathbf{x}$ and find that

$$x_1 x_2 x_3 = x_{10} x_{20} x_{30} e^{(e_{11} + e_{22} + e_{33})t}\tag{10}$$

Knowing that the sum of $e_{11} + e_{22} + e_{33}$ is zero of incompressible fluid, finally we get the trajectory line of a material element in the form of

$$x_1 x_2 x_3 = x_{10} x_{20} x_{30} = C\tag{11}$$

Assuming the flow is steady, the trajectory line of a material is the same as the streamline of the fluid. Thanks to that, Eq. (10) is also the form of streamline.

Secondly, we come back to the theme of defining collision rate between small drops in turbulent fluid. Without special statements, we still consider the fluid behavior in the moving reference. Taking the origin at the center of one drop with radius $r_1$, the other drops with radius $r_2$ are moving with the fluid along the streamlines. The collision would occur immediately when the two drops contact at the surface of sphere with radius R, which is the sum of $r_1$ and $r_2$, seeing in figure 2 (Saffman and Turner, 1956).

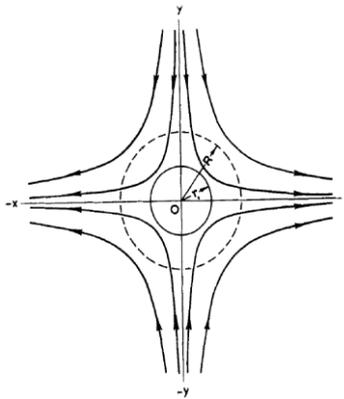

Figure 2. Streamlines of the relative motion of two dimension in the moving reference (Saffman and Turner, 1956)

However, figure 2 is not adequate to show the streamlines of the relative motion in three dimension space. According to Eq. (10), we plot another schematic diagram of the streamlines in three dimension space of the relative motion, seeing in figure 3. As from the figure, we can see that the streamlines of fluid composing a saddle surface to come across the surface of sphere with radius $R$.

If we use a plane parallel to *x-y* plane to cut out the sphere and the saddle surface, the equation (10) is reduced to

$$x_1 x_2 = C' \qquad (12)$$

since $x_3$ is constant at the cross section. (11) is thus a family of hyperbolical curves to show the streamlines in figure 2.

The next task is to determine the flux of fluid inwards across the surface of the sphere, for the collision rate of the 'fixed drop' at the origin is just the flux. No new derivation is done here and we take the result of Saffman and Turner's work directly. Supposing that the mean concentrations of two sizes of drops in a given population to be $n_1$ and $n_2$ per unit volume, we find the collision rate is

$$\frac{n_1 n_2}{2} \int \overline{|w_r|}\, ds \tag{13}$$

If we want to accomplish this integral, we should find the expression of $\overline{|w_r|}$, where $w_r$ means the radial component of the relative velocity, and the bar denotes a mean over many realization of the motion.

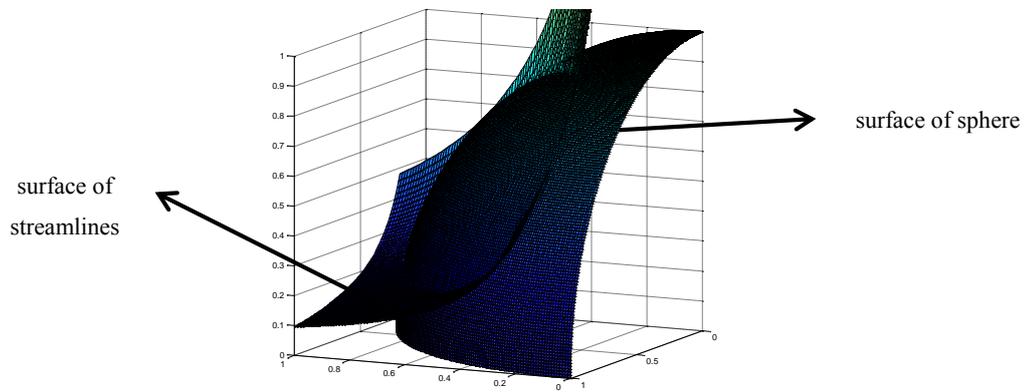

Figure 3. Streamlines of the relative motion of three dimension in the moving reference

In Saffman and Turner's work, the turbulence is assumed to be isotropic in the fixed reference($x'y'z'$, seeing in figure 1). The question comes to us that how the turbulent fluid is in the moving reference. To answer this question, we decide to analyze the force exerted on a material element with velocity $\boldsymbol{u}$ at position **x** at time t in the moving frame. Batchelor(1967) showed the fictitious body force per unit mass that acts on the material above in the moving frame is

$$-\boldsymbol{f}_0 - 2\boldsymbol{\Omega}' \times \boldsymbol{u} - \frac{d\boldsymbol{\Omega}'}{dt} \times \mathbf{x} - \boldsymbol{\Omega}' \times (\boldsymbol{\Omega}' \times \mathbf{x}) \tag{14}$$

The second term of equation (14) is the well-known coriolis acceleration which is always perpendicular to $\boldsymbol{u}$, hence it would not add the magnitude of the $\boldsymbol{u}$. In addition, $\boldsymbol{\Omega}'$ is not changing with time t in one small eddy, which leads to the third term of above equation to be zero. The term $-\boldsymbol{\Omega}' \times (\boldsymbol{\Omega}' \times \mathbf{x})$ can be rewritten to another form

$$-\mathbf{\Omega}' \times (\mathbf{\Omega}' \times \mathbf{x}) = \frac{1}{2}\nabla(\mathbf{\Omega}' \times \mathbf{x})^2 \tag{15}$$

As $\mathbf{\Omega}'$ is constant in one small eddy, $\mathbf{\Omega}' \times \mathbf{x}$ only depends on $\mathbf{x}$. So $\frac{1}{2}(\mathbf{\Omega}' \times \mathbf{x})^2$ is a conservation field and, the force produced by that is also a conservation force that could not transfer any kinetic energy to heat. In general, none of these three forces acting on the material element in the moving frame would devote it to the dissipated energy.

On the other hand, the first term $\mathbf{f}_0$ of expression (14) is the acceleration of origin $\mathbf{O}$ moving relative to the fixed frame, which is $\frac{D\mathbf{u}'(\mathbf{x}',t)}{Dt}$. For incompressible Newtonian fluid, the momentum equation of $\mathbf{u}'(\mathbf{x}',t)$ is

$$\frac{D\mathbf{u}'(\mathbf{x}',t)}{Dt} = -\frac{1}{\rho}\nabla P + \nu\nabla^2\mathbf{u}'(\mathbf{x}',t) \tag{16}$$

where $P$ is a modified pressure that contains the effect of gravity. Hence we can see that $\mathbf{f}_0$ is a dissipated term for it is affected by $\nu\nabla^2\mathbf{u}'(\mathbf{x}',t)$, which is the viscous acceleration. So we can image that the turbulent dissipation in the moving frame are dominated by the fictitious viscous term $\nu\nabla^2\mathbf{u}'(\mathbf{x}',t)$ and the real viscous term $\nu\nabla^2\mathbf{u}(\mathbf{x},t)$. Without further demonstration here, we assume the rate of energy dissipation $\epsilon_m$ in the moving frame is twice larger than $\epsilon$ in the fixed one.

Finally, we give the expression of collision rate to compare with the one provided in Saffman and Turner's paper. According to the fact that the small eddy where the collision of drops occur is isotropic, $\overline{|w_r|}$ is equal to $\overline{|w_1|}$ where $w_1$ is the radial velocity along the radius parallel to the x-axis. Since R is usually small compared with the length scale of the small eddy, $\overline{|w_1|} = R\left.\overline{\left|\frac{\partial w_1}{\partial x}\right|}\right|_{\mathbf{x}=0}$. Tennekes and Lumley(1972) showed the mean square of the velocity gradient is related to $\epsilon_m$ and $\nu$ through the expression $\overline{(\frac{\partial w_1}{\partial x})^2} = \frac{\epsilon_m}{15\nu}$. We now assume that $\frac{\partial w_1}{\partial x}$ is normally distributed as well as Saffman and Turner, then $\overline{\left|\frac{\partial w_1}{\partial x}\right|} = \sqrt{\frac{2\epsilon_m}{15\pi\nu}} = \sqrt{\frac{4\epsilon}{15\pi\nu}}$. The collision rate is thus

$$N = n_1 n_2 (r_1 + r_2)^3 \sqrt{\frac{16\pi\epsilon}{15\nu}} \tag{17}$$

And the collision rate provided by Saffman and Turner is

$$N_s = n_1 n_2 (r_1 + r_2)^3 \sqrt{\frac{8\pi\epsilon}{15\nu}} \tag{18}$$

At the end, we can find the collision rate given in our paper is related to Saffman and Turner's through the expression as below

$$N = \sqrt{2} N_s \tag{19}$$

Then, 2e can see from figure 4 that the collision rate obtained in our paper is a little bit higher than the one of Saffman and Turner.

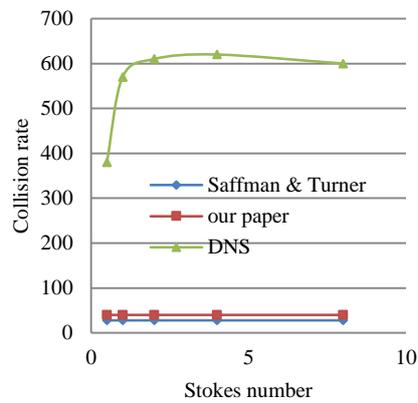

Figure 4. Average collision rate as a function of Stokes number

## 3. Conclusion

The flow near two close drops is considered to be not only uniform but also rotational. Hence a moving reference frame is built on a fixed drop to study the flux towards it. The trajectories of droplets moving near the fixed drop form a family of saddle surfaces. The collision rate of drops in this paper, which is more close to the simulation results got by DNS (Sundaram and Collins, 1997), is $\sqrt{2}$ times larger than the one of Saffman and Turner,.

**Reference**

Batchelor, G.K. An Introduction to Fluid Dynamics. UK, Cambridge; Cambridge University Press; 1967.

Grabowski, W.W., Wang, L.P. Growth of cloud droplets in a turbulent environment. Annual Review of Fluid

Mechanics, 2013, 45:293-324.

Marshall J.S, Li S.Q. Adhesive Particle Flow: A Discrete-element Approach. USA, New York: Cambridge University Press; 2014.

Saffman P.G., Turner J.S. On the collision of drops in turbulent clouds. Journal of Fluid Mechanics, 1956, 1:16-30.

Sundaram S., Collins L.R. Collision statistics in an isotropic particle-laden turbulent suspension .I. Direct numerical simulations. Journal of Fluid Mechanics, 1997, 335:75-109.

Tennekes H., Lumley J.L. A first course in turbulence. USA, MA, Cambridge; MIT Press; 1972.

Zhang H., Liu Q.F., Qin B.K., Bo H.L. Simulating particle collision process based on Monte Carlo method. Journal of Nuclear Science and Technology, 2015, 52(11): 1393-1401.

Zhou Y., Wexler A.S., Wang L.P. Modelling turbulent collision of bidisperse inertial particles. Journal of Fluid Mechanics 433, 77-104 (2001).